# *Electrochemically induced switching from antiferromagnetic spin-chain to frustrated spin-glass state in maple-leaf lattice $Na_2Mn_3O_7$*

*Corson Chao[1], Shivani Srivastava[2,3], Ming Lei[1], Bachu Sravan Kumar[1], Varun Kamboj[1], Hari Ramachandran[4], Zhelong Jiang[4,5], Anirudh Adavi[1], Kevin H. Stone[6], Mark Asta[2,3], Lilia S. Xie[7,*], Iwnetim I. Abate[1,*]*

1. *Department of Materials Science and Engineering, Massachusetts Institute of Technology, Cambridge, MA 02139, USA*

2. *Department of Materials Science & Engineering, University of California, Berkeley, California 97420, USA*

3. *Materials Sciences Division, Lawrence Berkeley National Laboratory, Berkeley, CA 94720, USA*

4. *Department of Materials Science and Engineering, Stanford University, 496 Lomita Mall, Stanford, CA 94305, USA.*

5. *Applied Energy Division, SLAC National Accelerator Laboratory, Menlo Park, California 94025, United States*

6. *Stanford Synchrotron Radiation Lightsource, SLAC National Accelerator Laboratory, Menlo Park, California 94025, United States*

7. *Department of Chemistry, University of California, Berkeley, California 97420, USA*

*[*] Corresponding-Author, E-mail: iabate@mit.edu and liliaxie@berkeley.edu*

## Abstract

We report the electrochemical tuning of magnetic properties in the $Na_2Mn_3O_7$ maple-leaf lattice (MLL) through ion deintercalation, revealing a switch from the 1D antiferromagnetic (AFM) spin-chain behavior of the S=3/2 MLL structure to frustrated magnetism spin-glass behavior. By utilizing Na deintercalation, we stabilize ferromagnetic (FM) short-range interactions within the original short-range AFM system, creating magnetic frustration within the system beyond that induced from the MLL geometrically frustrated structure, leading to a spin-glass state. Magnetic and structural analyses, combined with density functional theory (DFT) calculations, demonstrate the near-degeneracy between AFM and FM configurations in $Na_2Mn_3O_7$, suggesting that the altered lattice distortions and disorder introduced via deintercalation are responsible for the frustrated magnetism. Our findings provide a novel platform for studying low-dimensional magnetism, spin glass behavior, and potential applications in spintronics and computing technologies. This study represents the first observation of an induced spin glass state in MLL materials and is a rare example of electrochemically induced spin glass state, highlighting the critical role of ion intercalation in tuning magnetic interactions.

## Introduction

There is growing interest in low-dimensional magnetic systems, not only for advancing the understanding of quantum phenomena but also for their potential applications in quantum computing, spintronics, and magnetic sensing technologies.[1,2] Techniques such as ion doping, induced lattice strain, and electrostatic gating have been employed to fine-tune the magnetic properties of layered magnets, providing valuable insights into the mechanisms governing low-dimensional magnetism.[3–10] Among these methods, electrochemical ion intercalation stands out as a particularly effective approach, enabling real-time, controlled adjustments to magnetic properties for deeper investigation and potential technological applications.[3,11,12] Ion intercalation allows for the tuning of interlayer spacing, oxidation states, and exchange mechanisms within the material, resulting in a diverse range of magnetically ordered phases including antiferromagnetism, ferrimagnetism, ferromagnetism, spin-glass, and the modulation of magnetic ordering temperatures.[6,13–15]

In low dimensional magnetism, the maple leaf lattice (MLL) structure is a relatively unexplored area with much potential for studying the interplay of 1D and 2D magnetism due to the geometrically frustrated structure of the layered lattice.[16,17] The MLL can be derived from a triangular lattice with 1/7 of the sites removed in an ordered fashion. These ordered vacancies result in increased geometric frustration due to the lower magnetic connectivity and increased lattice distortions within the MLL,[16,18–22] which may give rise to distinctive physical phenomena, analogous to the properties that emerge in other solids with highly geometrically frustrated structures, such as Kagome lattices.[23–27] However, fundamental physical measurements of materials with the MLL structure remain limited due to the scarcity of material candidates with this configuration compared to triangular or Kagome lattices. Furthermore, other known materials with MLL structures often contain nonmagnetic impurities or extra magnetic ions between the layers, which can obscure the intrinsic properties of the MLL.[15,20] The manganese oxides $Na_2Mn_3O_7$ and $MgMn_3O_7{\bullet}3H_2O$,[18] which can be synthesized without impurities or extra magnetic ions between layers, are ideal platforms for studying the magnetic properties of MLL materials, and ion intercalation is a suitable way to tune the properties of these MLL materials. By de/intercalation of the sodium non-magnetic cations in varying amounts, the charge within the layered structure can be modulated, leading to switching of the oxidation state of the oxygen and manganese ions within the $[Mn_3O_7]^{2-}$ interlayers. This change of oxidation state leads to altered exchange interactions, such as in the degree of ferromagnetic double exchange mechanisms and antiferromagnetic superexchange, leading to the active tuning of the sample magnetic response.[28–33] Additionally, the layered MLL material $Na_2Mn_3O_7$ has been found to be highly reversible upon de/intercalation, possibly allowing for greater tunability and controlled observation of alterations to magnetic properties.[34,35]

Due to inherent geometric frustration in lattices such as MLL, it is difficult for the magnetic moments on the lattice to align in a low-energy configuration, often leading to degenerate ground states and varied disordered spin states at low temperatures. Quantifying and controlling the degree of disorder and frustration could potentially enable the engineering of frustrated spin systems and the discovery of even new phases. Materials with MLL structure have been found to form 1D antiferromagnetic spin-chain compounds and are also theorized to potentially host spin-liquid states.[15,18–20,20,36–38] Spin glasses, which harbor a multitude of ground-state degeneracies, are another form of frustrated spin state that have drawn considerable interest over the past few decades because they provide a platform to study non-equilibrium physics and can be used to transport angular momentum or spins for application in spintronics.[1,39] While more scalable and less prone to stray magnetic fields than its ferromagnetic counterpart, antiferromagnetic magnetoresistive random-access memory (MRAM) faces limitations in its responsiveness to external stimuli. The integration of spin glass states into antiferromagnetic materials has the potential to augment and stabilize antiferromagnetic memory switching,[39] leading to new mechanisms for switching antiferromagnets.

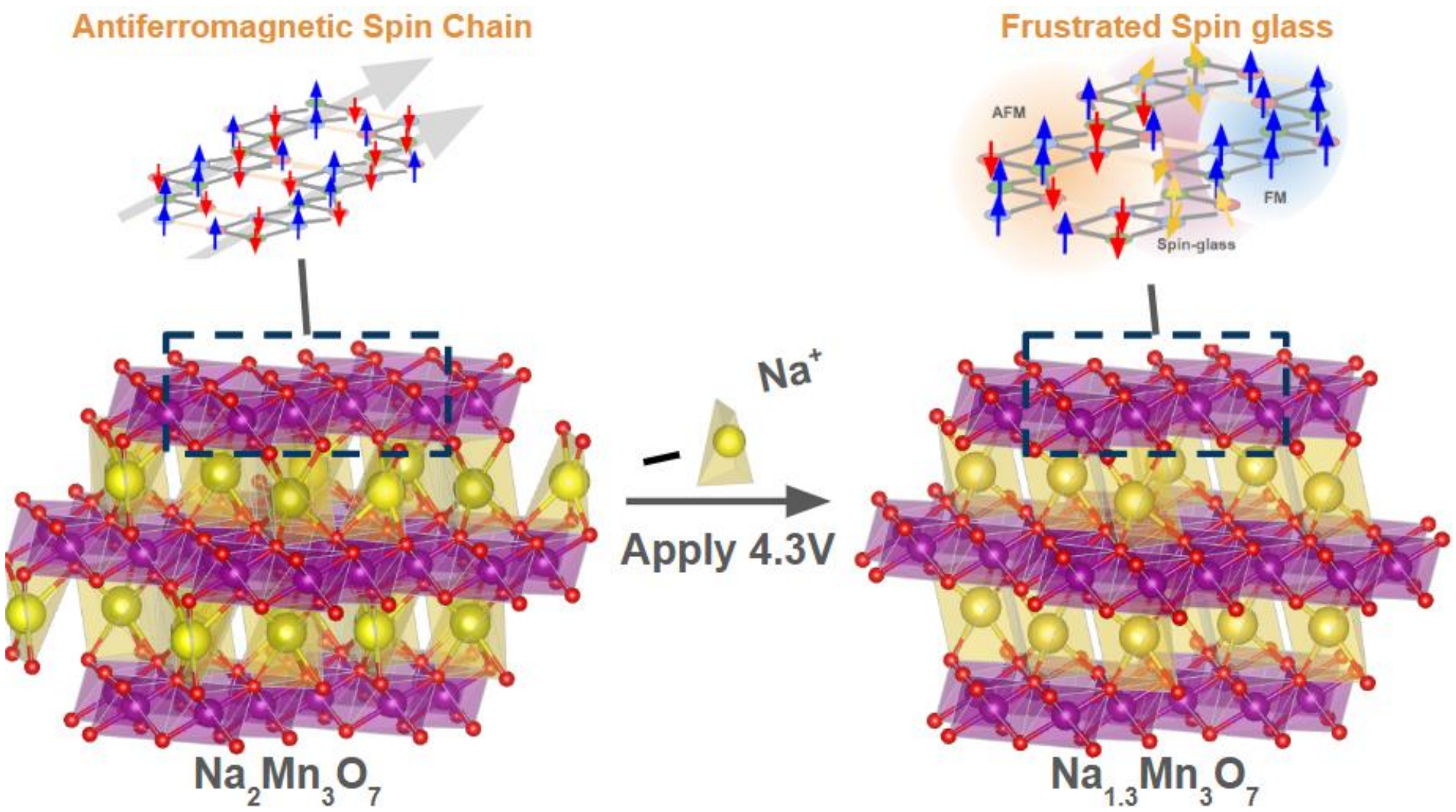

Figure 1: Electrochemically induced magnetic switching in maple-leaf lattice $Na_2Mn_3O_7$: From 1D antiferromagnetic spin-chain to frustrated spin-glass upon Na de-intercalation.

As shown in Figure 1, we seek to investigate the magnetic properties of $Na_2Mn_3O_7$ to elucidate whether the usage of ion intercalation can tune the magnetic properties and geometric frustration of the MLL. Combining AC and DC magnetization measurements and DFT calculations, we find that through the process of sodium de-intercalation, the 1D antiferromagnetic properties of $Na_2Mn_3O_7$ can be tuned to become spin-glass like in nature due to the creation of ferromagnetic regions within the material. To our knowledge this is the first report of the spin-glass state in materials with MLL, and a rare example of electrochemically induced spin-state.[40] Our DFT calculations suggest a minimal difference in energy between FM and AFM configuration (near-degeneracy). Presence of multiple possible near-degenerate orders,[41] in addition to the distorted lattice, are likely sources of the observed frustrated magnetism and spin-glass.

## Methods

$Na_2Mn_3O_7$ powder was prepared using a solid-state method. A stoichiometric mixture of $NaNO_3$ (J. T. Baker, A.C.S. Reagent) and $MnCO_3$ (Aldrich, ≥ 99.9%) powder was mixed with mortar and pestle for 30 mins. The collected mixture powder was then heated at 550 °C (5 °C/min) for 12 h in ambient conditions. The prepared powder sample was left to cool down naturally to 150 °C before it was transferred directly to an argon-filled glovebox (MBraun, $O_2$ and $H_2O$ ≤ 0.1 ppm), with air and moisture exposure minimized with the high temperature (150 °C) transfer. The charging of the sample was performed through electrochemical methods. The $Na_2Mn_3O_7$ powder was mixed with carbon black and polyvinylidene fluoride (PVDF) in an 80:10:10 ratio with N-methylpyrrolidone (NMP) solvent through the slurry process onto an aluminum current collector and dried through X process. A half-cell in coin cell configuration was constructed using sodium metal as the anode and the $Na_2Mn_3O_7$ mixture as the cathode. The half-cell is then charged beyond the voltage plateau at 4.2V to 4.3V where the sodium in the distorted octahedral site is removed creating $Na_xMn_3O_7$, x=~1.3 (Figure S1). The desodiated NMO cathode is then scraped off for magnetometry studies. The magnetization of pure $Na_2Mn_3O_7$ powder was compared to that of the scraped cathode containing aluminum, carbon black, and PVDF mass contributions with results showing the additional cathode components being non-magnetic (Figure S2). Synchrotron powder X-ray diffraction ($\lambda$ = 0.7292 Å) in capillary mode was used to determine the room temperature crystal structure in Figure S3.a. and was conducted at Stanford Synchrotron Radiation Lightsource (SSRL) on Beamline 2-1. Rietveld refinement of XRD patterns corresponding to $Na_2Mn_3O_7$ and de-sodiated $Na_2Mn_3O_7$ (i.e., $Na_{1.3}Mn_3O_7$) was performed with GSAS2 software and Topas Academic v7.[42]

The variable temperature XRD was collected using PANalytical X'Pert PRO with Cu Kα X-ray source in reflection mode using the cryostat cooling stage. All magnetometry experiments were performed on a Quantum Design Physical Property Measurement System (PPMS) DynaCool equipped with a 9T magnet. DC and AC magnetization were measured using the AC Measurement System (ACMS II) option. (An adapted cooling protocol was used to ensure the absence of negative trapped field.[43])

Computational studies were performed with density-functional theory (DFT). Specifically, lattice relaxation and total energy calculations were performed using the Projector-Augmented Wave (PAW) method,[44] as implemented in the Vienna Ab initio Simulation Package (VASP).[34,45] The accurate calculation is done by using Heyd-Scuseria-Ernzerhof (HSE) screened hybrid functional.[46] The PAW potentials employed in the calculations treated the following electronic states as valence: Na: $3s^1$, Mn: $4s^1$, $3p^6$, $3d^6$ and O: $2s^2$, $2p^4$. The Brillouin zone was sampled with a 4×4×4 Monkhorst–Pack k-point grid, employing Gaussian smearing of 0.05 eV. The plane-wave cut-off was 600 eV. Relaxation of ions and cell shape was performed with convergence criteria of $10^{-5}$ eV in total energy, and until the forces on the atoms reached a magnitude less than 0.01 eV $Å^{-1}$, respectively.

**Results and Discussions**

Rietveld refinement of the PXRD data revealed structural parameters for $Na_2Mn_3O_7$ consistent with previous literature reports and confirmed the absence of any phase impurities and the formation of phase pure NMO(Figure 3.a).[47–49] The atomic coordinates for both pristine and desodiated samples use the same value as reported by Song et al. (Table S1),[47] whereas the lattice parameters are refined based on the measured sample. In the desodiated sample, the occupancy for the Na2 site, with coordinate [0.3264, 0.8216, 0.0803] is set to 0.3. The peak shape was fitted with Thompson-Cox-Hastings pseudo-Voigt profile, with size and strain broadening. Additionally, Stephens' model for anisotropic broadening was included.[50]

A very good agreement is obtained between the measured and calculated XRD patterns for both $Na_2Mn_3O_7$ desodiated $Na_2Mn_3O_7$ with $P\bar{1}$ crystallographic symmetry identified for both $Na_2Mn_3O_7$ and $Na_{1.3}Mn_3O_7$ (Table S2).

$Na_2Mn_3O_7$ is composed of $[Mn_3O_7]^{2-}$ layers built up with edge-sharing $MnO_6$ octahedra.[35,36] The Na ions occupy two different sites between the layers, labeled Na1 and Na2, with trigonal prismatic and distorted octahedral coordination, respectively (Figure S3.e.). In addition, compared to a perfect triangular Mn lattice, 1/7 of the Mn ions are missing in an ordered fashion, giving rise to the characteristic maple-leaf lattice geometry, Figure S3.c. The loss of the 1/7 Mn ions results in three types of nearest neighbor interactions between remaining Mn ions as shown in Figure S3.b. The remaining Mn within the structure occupies three unique lattice positions within the $MnO_6$ octahedra in the $Mn^{4+}$ oxidation state as labelled in Figure S3.d. The Mn–Mn interlayer distance is much larger than the in–plane Mn–Mn distance (~5.7Å vs. 2.8 Å), resulting in weak interlayer coupling in $Na_2Mn_3O_7$. As a result, the magnetic behavior is expected to be quasi-two-dimensional in nature,[51] making NMO a promising platform for studying non-equilibrium dynamics in low-dimensional magnetism. We deintercalated Na2 ions from $Na_2Mn_3O_7$ to form $Na_{1.3}Mn_3O_7$ with negligible structural change, retaining the trigonal maple leaf lattice structure (Figure S1 & S3).

Previous studies on $Na_2Mn_3O_7$ have revealed complex magnetic behavior, characterized by the absence of long-range magnetic ordering or phase transitions below room temperature.[36,37] However, there is a discrepancy in the proposed magnetic models for this material. In this context, there is a strong motivation to understand the mechanism underlying the magnetic behavior of $Na_2Mn_3O_7$. One study suggests that $Na_2Mn_3O_7$ follows a one-dimensional (1D) Heisenberg spin-chain model, where the spin of each atom interacts with the spins of its immediate neighbors through a coupling constant, J.[36] Another study; however, supports a broader two-dimensional (2D) short-range spin-spin correlation model.[36,37] The discrepancy between the studies may stem from the presence of defects or impurities in the $Na_2Mn_3O_7$ samples in both reports, as evidenced by paramagnetic signatures in the magnetic susceptibility at low temperatures in both papers.[36,37]

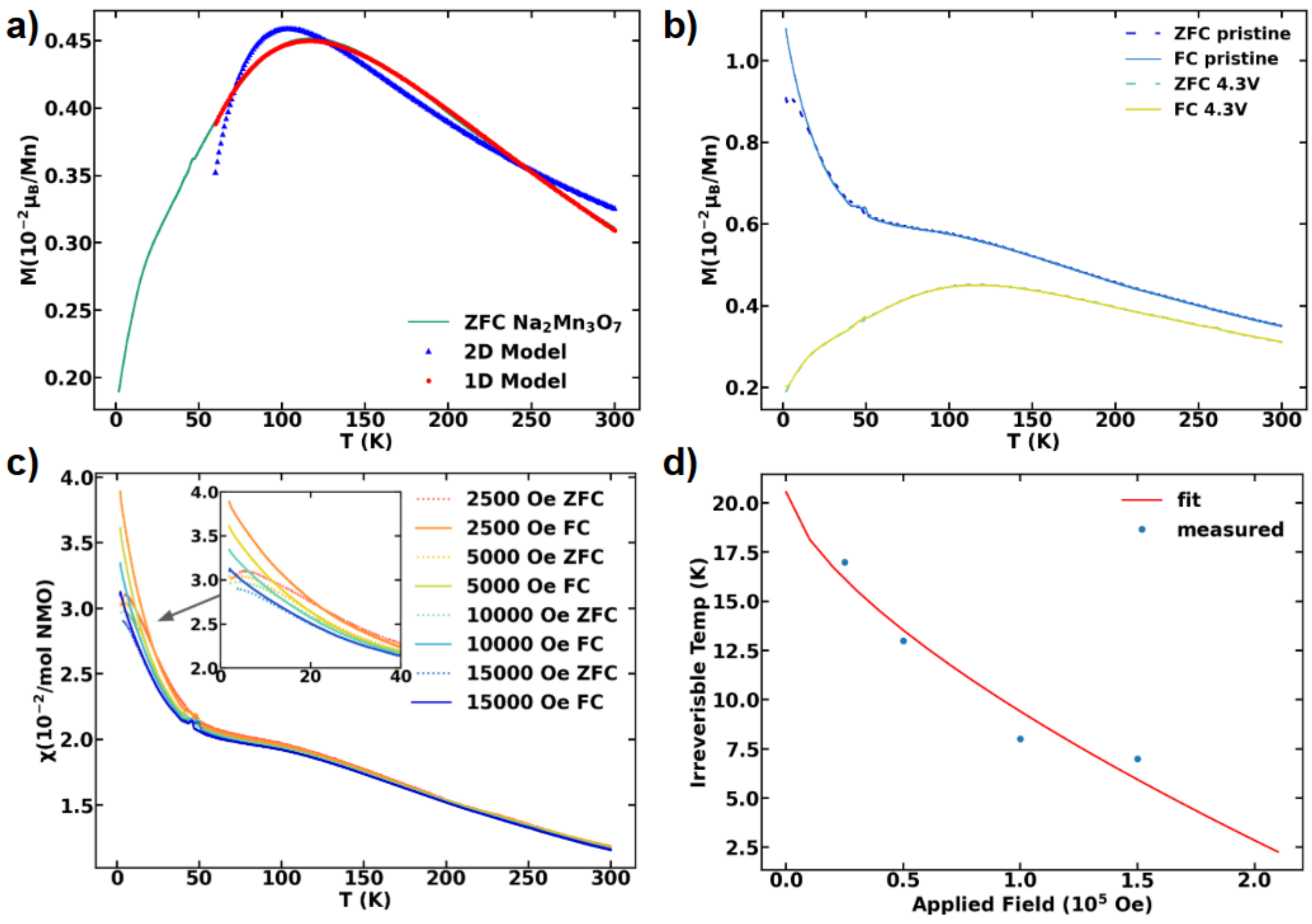

Figure 2. a) Zero Field-Cooled DC Magnetization of $Na_2Mn_3O_7$ against HTSE fits b) Comparative plot of ZFC and FC χ vs T for pristine $Na_2Mn_3O_7$(yellow) and de-intercalated $Na_{1.3}Mn_3O_7$ (green). c) Temperature-dependent zero field-cooled (ZFC) and field-cooled (FC) dc magnetization under applied magnetic fields of 2500, 5000, 10000, 15000 Oe for $Na_{1.3}Mn_3O_7$. The ZFC and FC data are represented by open and closed symbols, respectively d) Almeida-Thouless fit of bifurcation values for de-intercalated $Na_{1.3}Mn_3O_7$

Our measurement of the magnetic susceptibility of pristine $Na_2Mn_3O_7$ under ZFC and FC conditions at 5000 Oe, shows a cusp at ~ 120 K, with the susceptibility continuing to decrease as temperature is reduced (Figure 2.a.). This behavior suggests the presence of antiferromagnetic coupling in $Na_2Mn_3O_7$ with a broad peak ~ 120K indicative of short range spin correlations, consistent with prior literature.[36,37] Contrary to past reports, we find no Curie-like upturn at 10 K which had been previously attributed to impurities.[36,37] To further understand the nature of magnetic interactions, we applied the high temperature series expansion (HTSE) method to fit the temperature-dependent susceptibility curve with 2D and 1D short-range magnetic correlations models.[37,52,53] The ZFC susceptibility is found to be in better agreement with the 1D Heisenberg spin-chain model (Figure 2.a.), suggesting nearest-neighbor antiferromagnetic exchange coupling. The 1D model used is the 1D Heisenberg spin chain model with

$$\chi = \chi_0 + [Ng^2\mu_B^2 S(S+1)3k_BT]\frac{\sum_{l=1}^{7} 1+a_l(J/kT)^l}{\sum_{l=1}^{7} 1+b_l(J/kT)^l} \qquad (1)$$

where $a_1$ $to$ $a_7$ are 0.48166, 0.91348, 0.11791, 0.0054266, 0.00000065501, 0, 0 respectively and $b_1$ $to$ $b_7$ are 2.9834, 5.8136, 7.3541, 2.2637, 0.12999, 0, 0 respectively.[53] The 2D spin system, is represented by a power series given by

$$\chi = \chi_0 + [Ng^2\mu_B^2 S(S+1)3k_BT]\sum_{l=1}^{7} 1 + a_l(J/kT)^l \qquad (2)$$

where N is Avogadro's number, $k_B$ is the Boltzmann constant, $a_l$'s are coefficients, and $\chi_0$ represents the contributions from the temperature independent Van Vleck paramagnetic term and diamagnetic term due to atomic cores. The coefficient values used are as reported for the closest triangular lattice with $a_1$ $to$ $a_7$ being 4, 3.2, -2.186, 0.08, 3.45, and -3.99 respectively.[37,52]

As shown in Figure 2.b., upon deintercalation, the DC magnetic susceptibility exhibits paramagnetic, frustrated spin-glass as well as the original antiferromagnetic behavior. Alongside the upturn below 120 K, a clear paramagnetic signal becomes evident when the antiferromagnetic signal of the pristine compound is subtracted from the deintercalated magnetic susceptibility data, appearing linear when plotted against $1/\chi$ (Figure S4). In addition, bifurcation between the ZFC and FC states around 20 K is observed in the deintercalated sample, with the bifurcation temperature decreasing as the applied field increases (Figure 2c), consistent with behavior expected for a frustrated spin-glass state. Furthermore, the Almeida Thouless curve in Figure 2.d shows that the deintercalated sample follows the n=⅔ scaling behavior typical of spin-glasses, with irreversible temperature of 20.5K.[54]

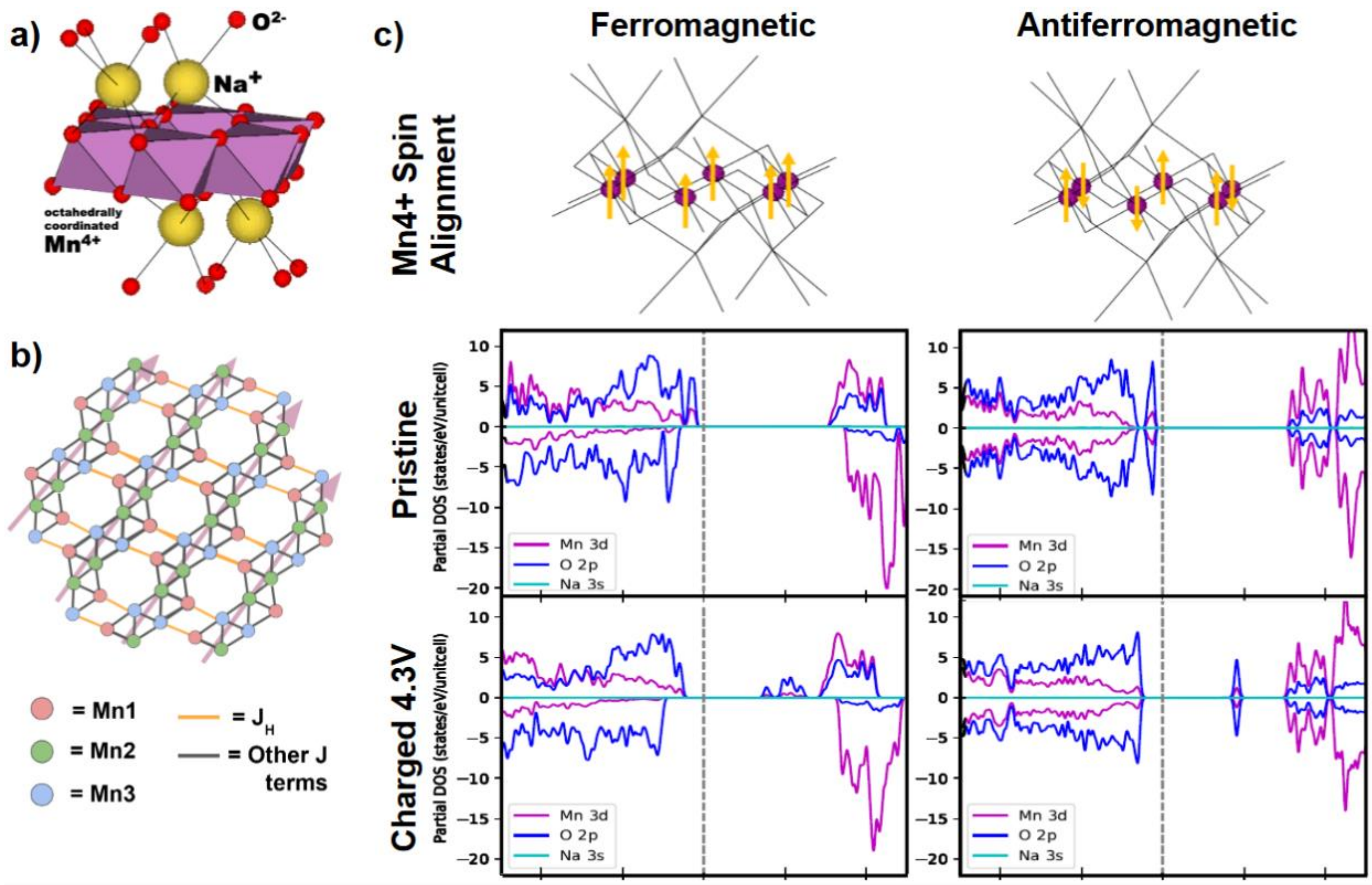

Figure 3. (a) Schematic diagram of ferromagnetic and antiferromagnetic ordering of $Na_2Mn_3O_7$. (b) model of 1D AFM ordering of pristine $Na_2Mn_3O_7$ and c) Partial density of states showing the orbital composition of different bands for FM and AFM $Na_2Mn_3O_7$ and $Na_{1.3}Mn_3O_7$.

The emergence of glassy magnetic behavior upon deintercalation can be attributed to the increase in magnetic disorder within the material. The schematic for ferromagnetic (FM) and antiferromagnetic (AFM) ordering in the layered $P\bar{1}$ $Na_2Mn_3O_7$ and $Na_{1.3}Mn_3O_7$ structures are shown in Figure 3.a.c with the calculated electronic band structure and orbital-decomposed density of states presented in Figure 3.c. For both materials, the bands are predominantly composed of hybrid O *p* and Mn *d* states at the valence band maximum (VBM) for both AFM and FM configurations. Our DFT calculations for $Na_2Mn_3O_7$ suggest that AFM is relatively more stable than FM in the pristine form by 2.75 meV/f.u. (10 times less than $k_BT$ *at room temperature*). Upon deintercalation, the FM form is calculated to be more stable than the AFM by 177 meV/f.u.. The spin stabilization during deintercalation is analyzed by comparing the energy stabilization of AFM and FM spin states in $Na_2Mn_3O_7$ and $Na_{1.3}Mn_3O_7$. By isolating the spin contribution from the total energy change—after accounting for the chemical energy difference between the pristine and deintercalated states—it is observed that the AFM spin state stabilizes by 148.73 meV, while the FM spin state stabilizes by 332.827 meV upon deintercalation (Table S3).

Using Rietveld refinement, we additionally performed atomic coordinate refinement of the $Na_2Mn_3O_7$ MLL structure, in which the distance of one Mn-Mn bond was found to be ~3 Å (referred as $J_h$ interaction) while the other two are ~2.7 Å, which is the characteristic behavior expected for 1D antiferromagnetic spin chain interactions, where the short-range spin order is stabilized by strong lattice distortions within the structure (Table S4). These distortions weaken the $J_h$ interaction, allowing spin correlations to favor a 1D antiferromagnetic spin-spin correlations as shown in Figure 3.b and in literature

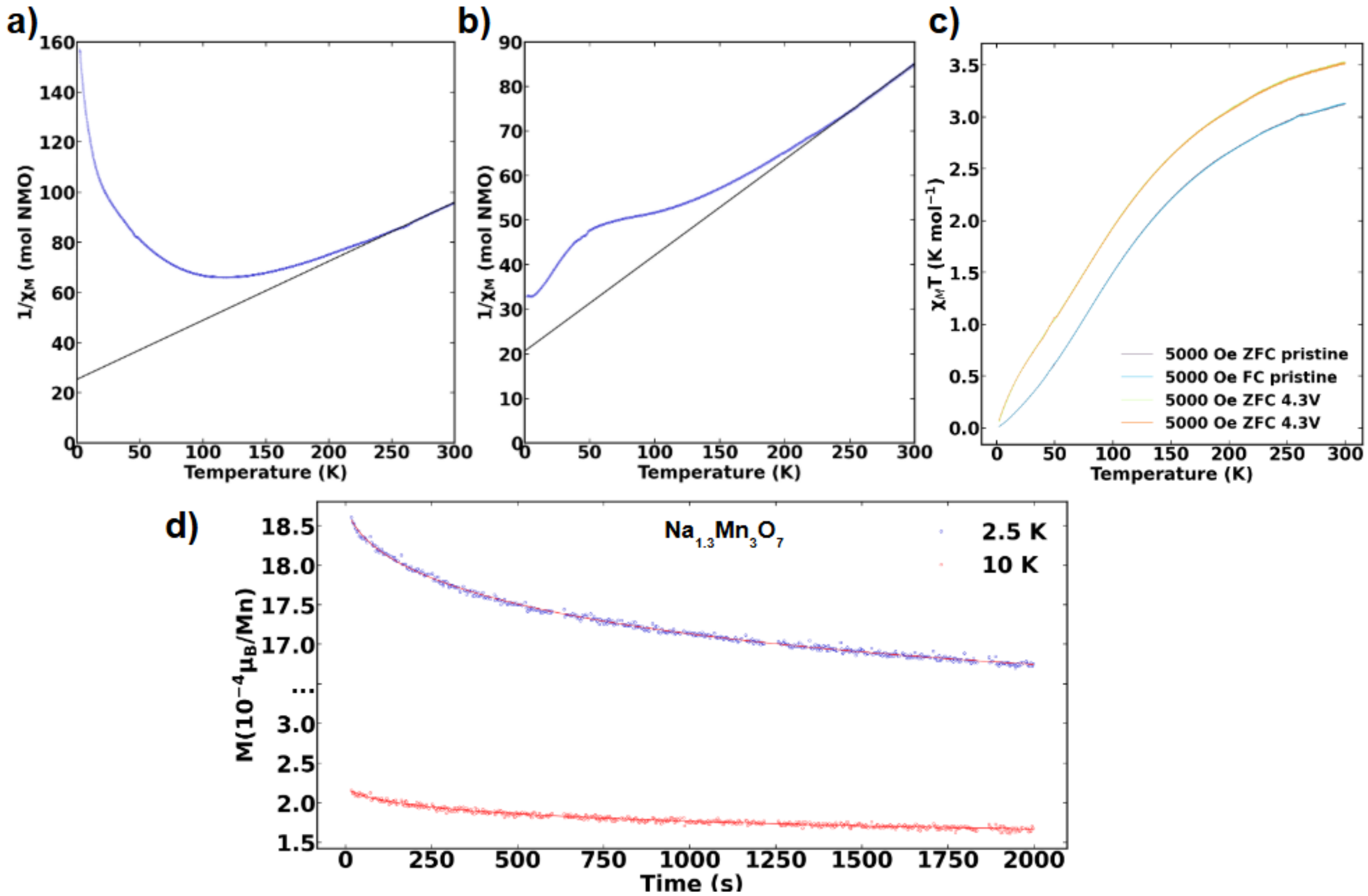

Figure 4 (a.b) Inverse susceptibility vs. temperature data with a 5000 Oe field, fitted to the Curie–Weiss law between ~250 and 300 K (a) $Na_2Mn_3O_7$ and (b) $Na_{1.3}Mn_3O_7$ . (c) $\chi_M T$ vs. temperature with pristine representing $Na_2Mn_3O_7$ and 4.3V representing the charged $Na_{1.3}Mn_3O_7$ (d) Thermoremanent magnetization of $Na_{1.3}Mn_3O_7$

The magnetic behavior of $Na_2Mn_3O_7$ and $Na_{1.3}Mn_3O_7$ at higher temperatures (250 K – 300 K) is characterized by a Curie–Weiss temperature, $\theta_{CW}$, of –122.33 K and -81.93 K and a Curie constant, *C,* of 4.41 and 4.45 emu K $Oe^{-1}$ $mol^{-1}$, respectively (Figure 4.a.b). The negative value and appreciable magnitude of $\theta_{CW}$ is consistent with the prevalence of antiferromagnetic interactions in this temperature regime, possibly resulting from Mn–Mn direct exchange interactions.[3] The reduction in the magnitude of $\theta_{CW}$ upon deintercalation indicates the shift from antiferromagnetic to paramagnetic behavior. In addition, due to the distorted octahedral geometry around each Mn center, the Mn–O–Mn bond angles deviate considerably from 90°.[36] As a result, according to Goodenough–Kanamori rules, ferromagnetic Mn–O–Mn superexchange interactions are not expected, but antiferromagnetic superexchange interactions are possible.[55,56] The $\theta_{CW}$ of $Na_{1.3}Mn_3O_7$ is significantly lower in magnitude and closer to zero compared to the antiferromagnetic $Na_2Mn_3O_7$, suggesting the increased paramagnetic or mixed AFM/FM interactions upon deintercalation. Moreover, the value of the *C* is significantly lower than the theoretical value of 5.63 emu K $Oe^{-1}$ $mol^{-1}$ for three $Mn^{4+}$ ions (assuming $g = 2$ and $S = 3/2$). Overall, deviation from ideal Curie–Weiss behavior suggests that short-range interactions may affect the magnetization significantly, even up to 300 K. Additionally, the monotonically increasing value of $\chi_M T$ up to 300 K (Figure 3.c.) suggests that a temperature-independent term may also contribute to the observed moment at high temperatures. The different $\theta_{CW}$ and *C* compared to previous reports (–192 and –152 K for $\theta_{CW}$; 5.77 and 5.7 emu K $Oe^{-1}$ $mol^{-1}$ for *C*)[3,5] likely arise from the phase homogeneity of our samples, which could be a result of our sample preparation procedure.

In Figure 2.b, $Na_{1.3}Mn_3O_7$ exhibits a significant bifurcation of ZFC and FC below 20 K, suggesting potential spin-glass behavior. A distinct characteristic of spin glass behavior is the frequency dependence of the real part ($\chi'$) of the AC magnetic susceptibility. To investigate the presence of a spin glass state, we attempted temperature-dependent ac susceptibility measurements at various frequencies with an applied ac magnetic field of 15 Oe (both with and without a DC field). However, we found that the measured value of $\chi'$ (~$10^{-7}$ emu) fell below the resolution of the PPMS instrument (~$10^{-6}$ emu), rendering the data unreliable. Consequently, we opted for an alternative method to probe spin-glass behavior, utilizing time-dependent thermoremanent magnetization (TRM). We first cooled to either 2.5 K or 10 K in an applied field of 30 kOe. Once the target temperature was reached, the field was maintained for another 600 s before turning off. The relaxation was then recorded for > 30 minutes, Figure 5.d,e.  A slow decay of TRM was observed for $Na_{1.3}Mn_3O_7$ while no TRM was observed for $Na_2Mn_3O_7$ (Figure S5, Table S5). The slow decay in $Na_{1.3}Mn_3O_7$ could be a characteristic of spin-glass behavior which we analyzed further by fitting a modified stretched exponential function:

$$M(t) = M_0 + M_g exp\left[\left(-\frac{t}{\tau}\right)^{(1-n)}\right] \qquad (3)$$

where $M_0$ and $M_g$ are the intrinsic and glassy components of the moment, respectively, $\tau$ is the average relaxation time, and $n$ is the time stretch exponent. The values of $\tau$ obtained for $Na_{1.3}Mn_3O_7$ are around 1175 and 1185 seconds at both 2 K and 10 K (Table 2), which indicate a clear evolution of the magnetization over time. The values of $n$, 0.47(1) and 0.52(1) at 2.5 K and 10 K, respectively, are typical for spin glass systems.[57]

Table 2. Parameters from fitting the time-dependent remnant magnetization to a modified stretched exponential function for $Na_{1.3}Mn_3O_7$.

| T (K) | $M_0$ ($\mu_B/Mn$) | $M_g$ ($\mu_B/Mn$) | $\tau$ (s) | $n$ |
|---|---|---|---|---|
| 2.5 | 0.001598(7) | –0.000288(9) | 1178(79) | 0.47(2) |
| 10 | 0.000141(7) | –0.00008(10) | 1199(327) | 0.52(6) |

The geometrically frustrated maple leaf lattice and lattice disorder (due to unequal Mn–Mn distances and Mn–O–Mn angles) are ingredients for spin glass freezing. The frustration protects the ergodicity of the system until it reaches the spin glass transition where a metastable phase is formed. In addition, as previously mentioned, our DFT calculation suggests that the energy difference between AFM and FM in $Na_2Mn_3O_7$ is only 2.75 meV/f.u. (10 times less than $k_BT$ *at room temperature*) while the de-intercalated form has FM as more stable by 177 meV/f.u.. This relatively small energy difference indicates a near-degeneracy between ferromagnetic and antiferromagnetic order (Figure 3). As reported in previous studies, such near-degeneracy can be a signature of competing super-exchange and direct exchange interactions in the maple leaf lattice.[5,9] Since we do not expect this material to show itinerant magnetism due to localized Mn moments, the total energy method for calculating exchange constants should give reasonable results.[58,59] As a very crude approximation, the difference in energy of FM and AFM configurations divided by the number of nearest neighbors at each site should provide us an estimate of exchange constants. In this paper, we have not aimed at computing quantitative values for these exchange constants which will require sampling multiple magnetic configurations and larger supercells. The presence of multiple possible near-degenerate competing ordering of moments which cannot all be satisfied simultaneously have previously been shown to lead towards the spin-glass behavior, and we anticipate a similar origin of spin-glass state in this system.[60]

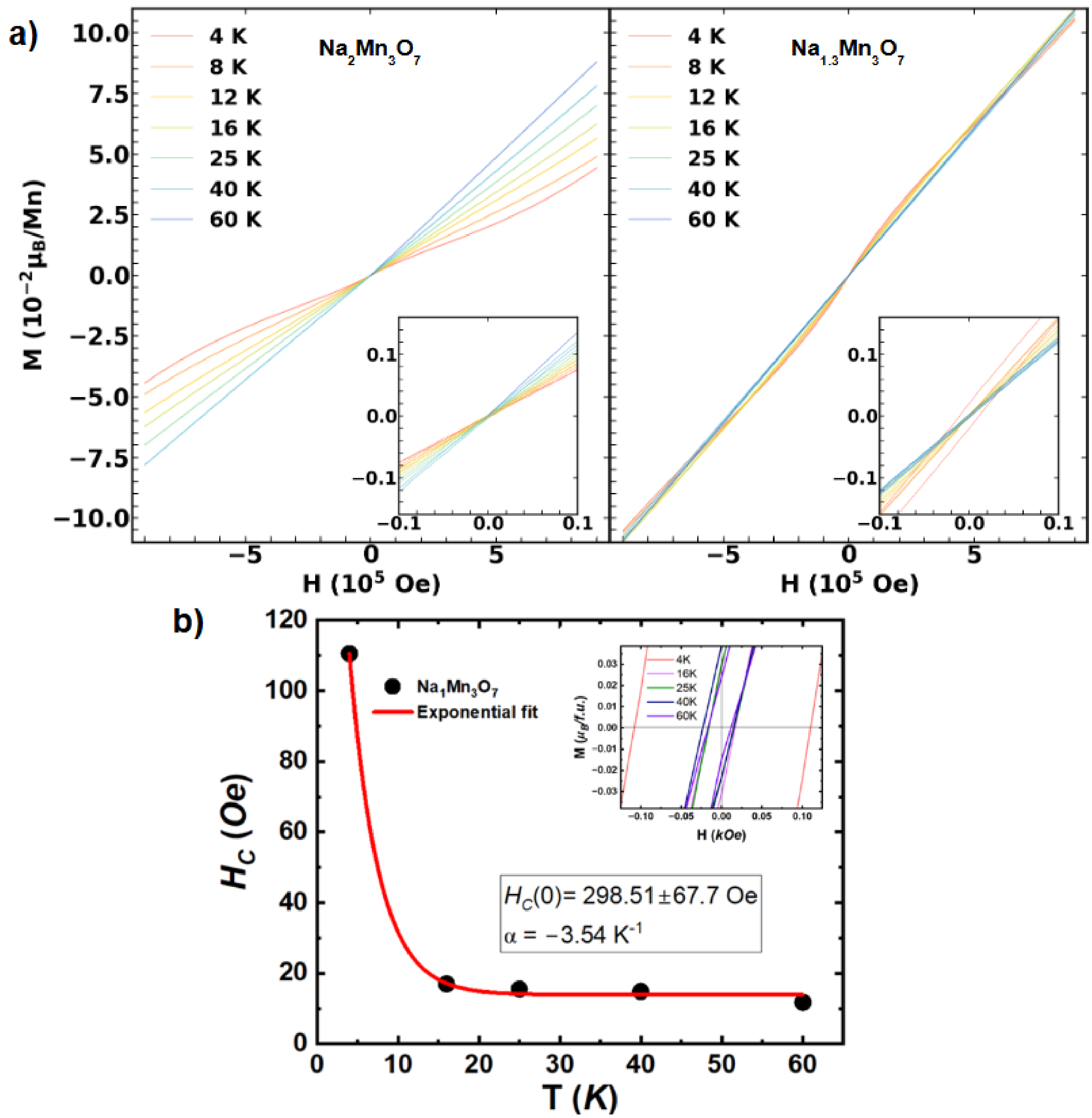

Figure 6 (a) Field-dependent dc magnetization measured at various temperatures between –90 and 90 kOe for $Na_2Mn_3O_7$ (left) and $Na_{1.3}Mn_3O_7$ (right), respectively. The insets are enlarged views of the hysteresis in field-dependent dc magnetization measured at -0.1 to 0.1T, or -1000 to 1000 Oe. (b) Coercive field, $H_C$, from dc magnetization as a function of temperature (T) for $Na_{1.3}Mn_3O_7$. The red line shows the fits using equation 4 predicted for spin glass systems. The inset shows an enlarged view of dc magnetization measured at various temperatures in the range 0 to 0.125 kOe.

Isothermal magnetization experiments revealed a lack of saturation in the magnetization values even at the largest applied fields of ± 90 kOe for both $Na_2Mn_3O_7$ and $Na_{1.3}Mn_3O_7$ (Figure 6.a). The maximum magnetizations of ~0.15 $\mu_B$/formula unit at +9 kOe is significantly less than expected for three $Mn^{4+}$ ions in the paramagnetic regime (6.70 $\mu_B$), consistent with antiferromagnetic behavior and spin frustration from the maple leaf lattice. The magnetization of $Na_2Mn_3O_7$ shows linear behavior with minimal hysteresis across temperatures (Figure 6.a., left) with the slope of the plot, $\chi$, decreasing with decreasing temperature as the material displays increasing antiferromagnetic response. The slight curvature at low temperatures can be potentially attributed to spin canting.[37] The $Na_{1.3}Mn_3O_7$ displays similar $\chi$ across temperature ranges, implying a lack of long range order, but it shows a slight hysteresis with low retentivity. At 4K, we also find an emergent hysteresis loop of about 110 Oe within the material (Figure 6.a., right) with the hysteresis decreasing with increasing temperature, disappearing completely by 8 K to yield a linear magnetization vs. field response. Both spin-glasses as well as superparamagnetic systems are characterized by a narrow M(H) hysteresis loop without saturation with non-zero coercivity and remanence below $T_C$.[61,62] To elucidate between the nature of magnetic correlation upon deintercalation in $Na_{1.3}Mn_3O_7$ and discriminate between the spin-glasses and superparamagnetic phase, we performed a study of the temperature variation of coercivity. The M(H) hysteresis loops were measured at various temperatures and shown in Figure 6.a for $Na_{1.3}Mn_3O_7$. The extracted coercive field ($H_C$) determined from M(H) loops for each temperature is plotted in Figure 6.b. We find that the coercivity of $Na_{1.3}Mn_3O_7$ decays exponentially with increasing temperature obeying the empirical model predicted for spin glass systems below the transition temperature Tc:[62,63]

$$H_C(T)=H_C(0)\, e^{-\gamma T} \qquad (4)$$

Where $H_C$ (T) is the coercive field at 0 K, and $\gamma$ is a fitting parameter. We performed the least squares fit (red solid line Figure 6.b.) using equation 4 to obtain the fitting parameters $H_C(0)$ = (298.51± 67.7) Oe, and $\gamma$ = (3.54 ± 0.72) $K^{-1}$. The exponentially decaying behavior of $H_C$ with temperature has been shown as a key signature of spin glass state and cluster spin glasses.[62] Exchange bias is then utilized to find the presence of ferromagnetic states within the deintercalated, spin-glass structure.

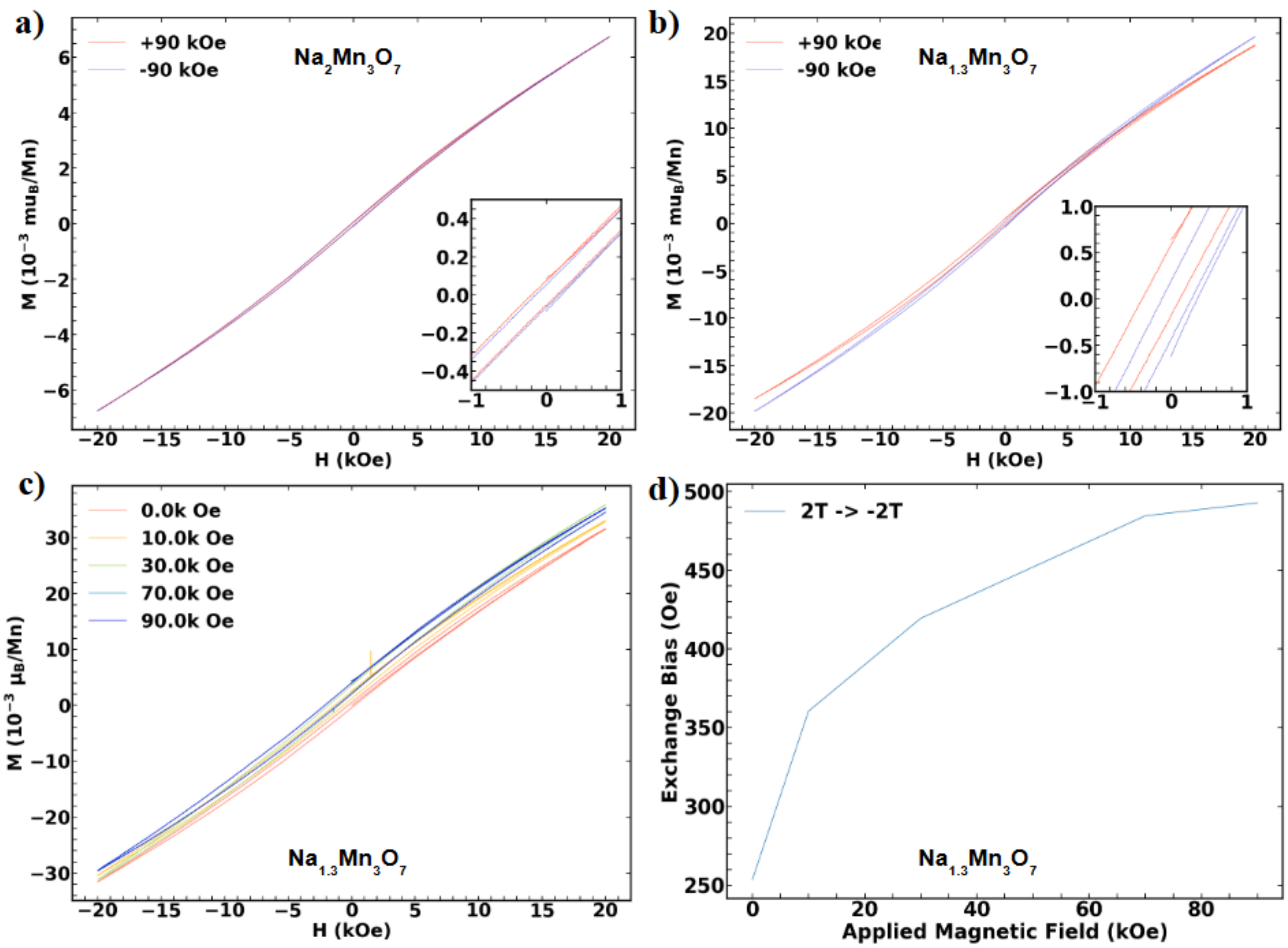

Figure 7 The (a) and (b) Field-cooled dc magnetization measured at 2.5 K between –90 and 90 kOe for $Na_2Mn_3O_7$ and $Na_{1.3}Mn_3O_7$ respectively. The insets are enlarged views of the exchange bias of the hysteresis loop for both systems (c) Exchange bias with field cooling of different applied magnetization for $Na_{1.3}Mn_3O_7$ (d) Change in exchange bias field with applied magnetic field for $Na_{1.3}Mn_3O_7$

Through exchange bias studies, we find pristine $Na_2Mn_3O_7$ to have no exchange bias, indicating the absence of ferromagnetic components in the material (Figure 7.a.). In contrast, the de-intercalated $Na_{1.3}Mn_3O_7$ shows a shift in the hysteresis curve when field cooled in 9T vs -9T, suggesting the presence of ferromagnetic layers that pin the magnetization at interfaces with either antiferromagnetic or spin-glass layers (Figure 7.b.).[64] By plotting the exchange bias at different magnitudes of applied field (Figure 7.c.d.), the exchange bias shows a logarithmic-like tapering of the exchange bias field, indicative of the saturation of the FM interfaces, which can be further used to estimate the FM magnetization, and layer thickness and size.[64,65] In future studies, dark field and Fresnel imaging can be used to further image the FM and AFM regions visually to confirm the spin-glass regions and phase coexistence predicted through DFT and exchange bias studies.[66] Temperature dependent heat capacity measurements can also be used to find if the intercalation can lead to phase segregation at low temperature as the material displays FM and AFM characteristics rather than only paramagnetic.[67] Collectively, these additional techniques could provide a next step to characterizing the structural and magnetic complexity of electrochemical tuning of magnetization, offering deeper insight into the mechanisms governing magnetic frustration and spin-glass formation within low-dimensional systems.

**Conclusion**

We have demonstrated electrochemical sodium ion deintercalation transforms $Na_2Mn_3O_7$ from a quasi-1D antiferromagnetic (AFM) spin-chain state to a frustrated spin-glass state, driven by ferromagnetic (FM) interactions and the geometric frustration of the maple-leaf lattice (MLL). Key signatures of spin-glass behavior, such as ZFC-FC bifurcation and slow relaxation, are observed in $Na_{1.3}Mn_3O_7$, marking the first demonstration of an induced spin glass state in MLL materials. Our work demonstrates how electrochemical de-intercalation can be used to induce and modulate exotic magnetic states making NMO an ideal platform to study the rich and complex interplay between structural distortions, intercalation-driven lattice modulation, and the emergence of novel spin-glass. Further work could study the coupling between electronic, crystal, and magnetic structures in such ordered-vacancy system with different vacancy concentration, transition metal types, and electrochemical de-intercalation towards new pathways for designing reconfigurable spin systems with applications in spintronics and quantum information technologies.

**Author Contributions**

I. I. A. conceived the project; I. I. A. and L. X. supervised the experiments; I. I. A. and M. A. supervised the theoretical calculations; C. C. and H. R. synthesized the material. C.C., K.S, A.A,. and H. R. performed XRD. H. R., B. S. K. and Z. J. performed the Reitveld refinement. ; C. C., I. I. A. and L. S. X. performed the magnetization measurements and analysis. S. S. and M.L. performed the DFT calculations. C.C, L. S. X and I. I. A. wrote the manuscript with inputs from all authors. I. I. A. directed the overall research.

**Conflicts of interest**

No conflict of interest to declare.

**Acknowledgments**

The authors acknowledge Kwabena Bediako, Zhizhi Kong, and Das Pemmaraju for insightful discussions and Hugh Smith for preparing samples for XRD. This material is based upon work supported by MIT Startup Fund. The authors also acknowledge the William Chueh Lab and the Kwabena Bediako Lab where some of the initial synthesis are performed. L.S.X. acknowledges support from the Arnold and Mabel Beckman Foundation (Award No. 51532) and L'Oréal USA (Award No. 52025) for postdoctoral fellowships. H.R. acknowledges funding support from the Knight-Hennessey Scholarship at Stanford University. S.S and M.A acknowledge funding support from the U.S. Department of Energy, Office of Science, Office of Basic Energy Sciences, Materials Sciences and Engineering Division, under Contract No. DE-AC02-05-CH11231 (Materials Project program KC23MP). Use of the Stanford Synchrotron

Radiation Lightsource, SLAC National Accelerator Laboratory, is supported by the U.S. Department of Energy, Office of Science, Office of Basic Energy Sciences under Contract No. DE-AC02-76SF00515. The DFT calculations used computational resources from Extreme Science and Engineering Discovery Environment (XSEDE), which is supported by the National Science Foundation under Grant No. ACI-1548562.

# Supplementary

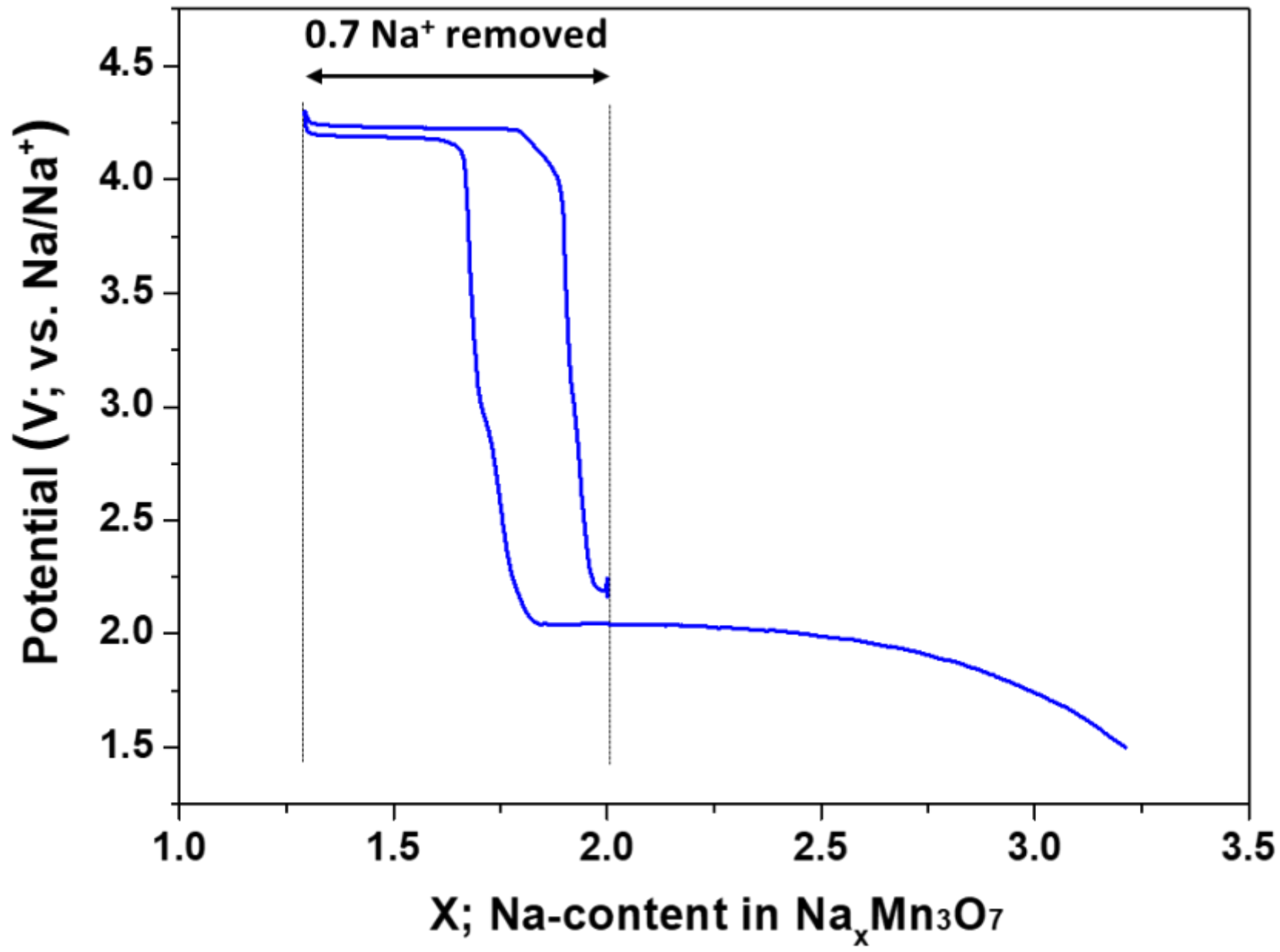

Figure S1: Charging profile of $Na_xMn_3O_7$ showing the plateau of distorted octahedral Na removal upon deintercalation until x=1.3, where Na1 begins to be removed, as shown through the rising voltage potential

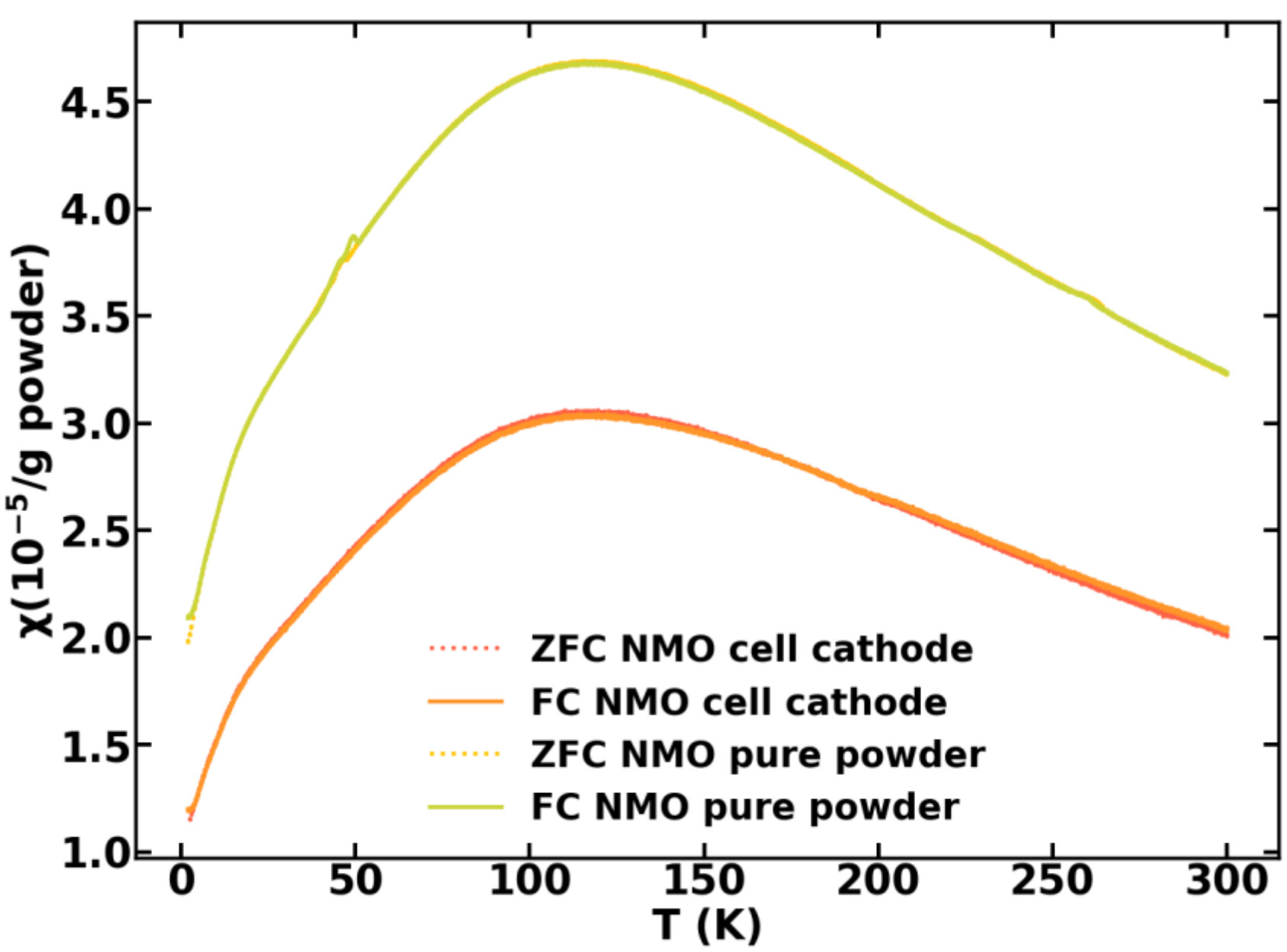

Figure S2: Zero-Field Cooled and Field-Cooled magnetization across temperatures for the pure synthesized $Na_2Mn_3O_7$ powder and the scraped cathode from the electrochemical cell, with around 30% mass consisting of non-magnetic carbon black, PVDF, and aluminum. The applied field was 5000 Oe for both the cell and powder.

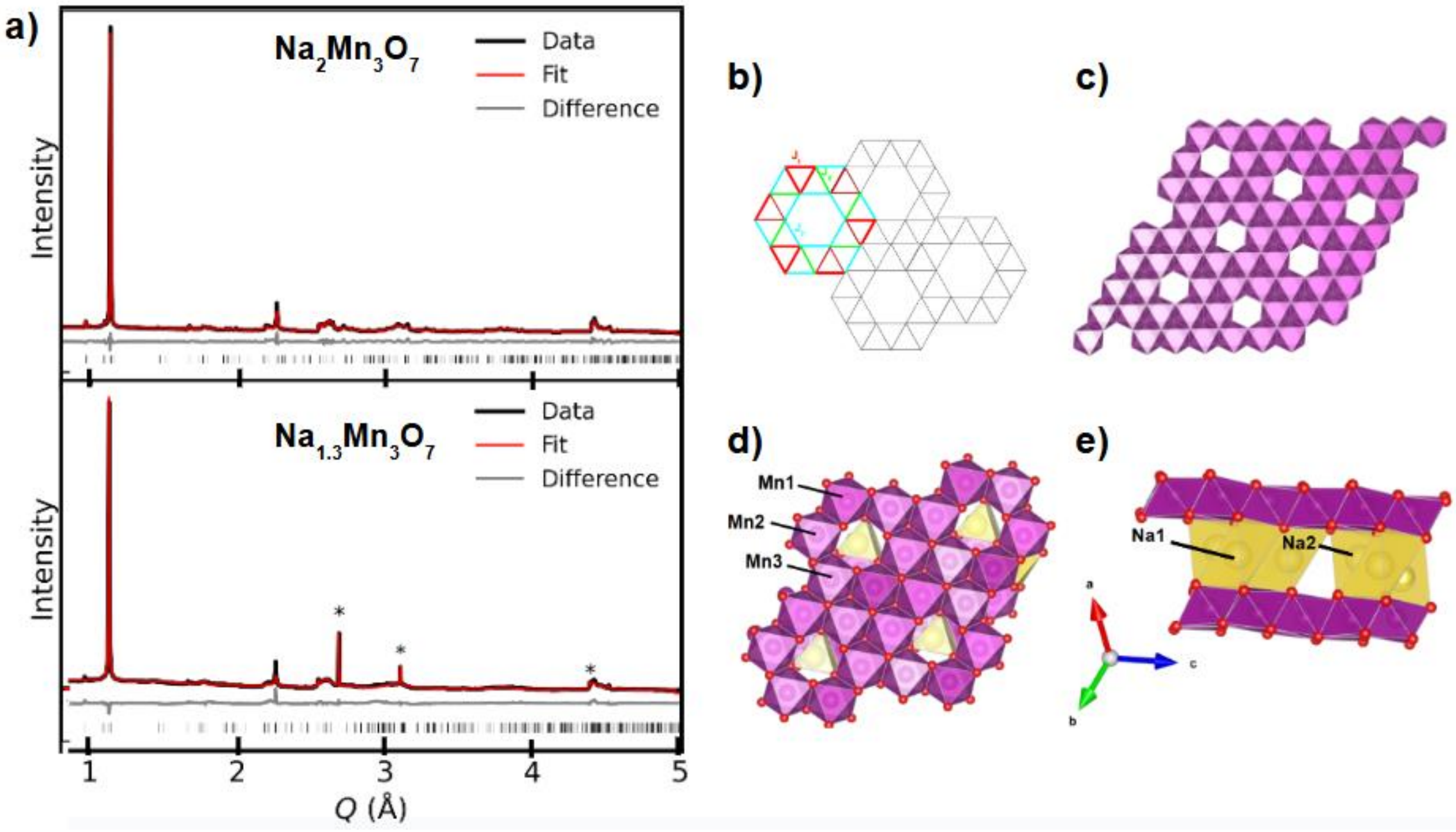

Figure S3 (a) Room-temperature X-ray diffraction pattern and Rietveld refinement on polycrystalline $Na_2Mn_3O_7$ powder. XRD was collected at SSRL X-ray source on Beamline 2-1 (λ = 0.7292 Å) In the electrochemically deintercalated $Na_{1.3}Mn_3O_7$, three additional sharp peaks are found at d = 2.33732, 2.02418, 1.43131, labelled "*", corresponding to the aluminum current collectors used in the electrochemical cycling process [68]. Small deviations of peak positions and intensities were found between measured and fitted patterns[47], but the deviations were not significant with no major impurity peaks present. The minor shifts may be attributed to synthesis variations in the poorly crystalline triclinic structure. (b) Schematic of the three exchange coupling constants of varying strength within the MLL lattice (c) Maple-leaf lattice (d, e) Schematic of top and side view of $Na_2Mn_3O_7$ crystal structure with the unique lattice sites for Mn and Na ions labelled. The two-dimensional layered arrangement of $MnO_6$ polyhedral sheet is separated by the nonmagnetic Na (yellow) layer.

Table S1: Utilized crystal structure of $Na_2Mn_3O_7$ with atomic coordinates taken without change from ref. ICDD PDF# 04-012-8515 [47] for Rietveld comparison. $Na_xMn_3O_7$ is a poorly crystalline triclinic structure so sample variations are expected

| $Na_xMn_3O_7$ Space Group: $P\bar{1}$ | | | | | | | |
|---|---|---|---|---|---|---|---|
| No. | Atom | Label | x | y | z | Occ. | No. |
| 1 | *Mn* | Mn1 | 0.0730 | 0.0737 | 0.2236 | 1 | 1 |
| 2 | *Mn* | Mn2 | 0.3632 | 0.3657 | 0.0860 | 1 | 2 |
| 3 | *Mn* | Mn3 | 0.2238 | 0.2250 | 0.6455 | 1 | 3 |
| 4 | *Na* | Na1 | 0.2440 | 0.6985 | 0.5059 | 1 | 4 |
| 5 | *Na* | Na2 | 0.3264 | 0.8216 | 0.0803 | (x-1) | 5 |
| 6 | *O* | O1 | 0.3883 | 0.2072 | 0.2380 | 1 | 6 |
| 7 | *O* | O2 | 0.2678 | 0.0873 | 0.8341 | 1 | 7 |
| 8 | *O* | O3 | 0.0284 | 0.2123 | 0.0306 | 1 | 8 |
| 9 | *O* | O4 | 0.1251 | 0.3235 | 0.4401 | 1 | 9 |
| 10 | *O* | O5 | 0.6912 | 0.5071 | 0.1136 | 1 | 10 |
| 11 | *O* | O6 | 0.1167 | 0.9384 | 0.4111 | 1 | 11 |
| 12 | *O* | O7 | 0.4531 | 0.6368 | 0.2944 | 1 | 12 |

Table S2, lattice parameters from Rietveld refinement for Na2Mn3O7 and Na1.3Mn3O7, space group P-1, atomic coordinates are taken without change from ref. ICDD PDF# 04-012-8515 [47]

| | a (Å) | b (Å) | c (Å) | α (°) | β (°) | γ (°) |
|---|---|---|---|---|---|---|
| Na2Mn3O7 | 6.5412(9) | 6.9071(9) | 7.5245(5) | 108.518(15) | 104.018(16) | 111.764(4) |
| Na1.3Mn3O7 | 6.5745(20) | 6.8896(19) | 7.5391(9) | 108.010(22) | 104.517(26) | 111.690(10) |

Table S3: Table of energy levels (eV) for the non-magnetic, AFM, and FM spin DFT for $Na_2Mn_3O_7$ and $Na_{1.3}Mn_3O_7$ relative to unbonded atoms as 0 eV for the comparison of stabilization energies due to spin orientations. The energy difference in non-magnetic deintercalation is due to the chemical stabilization from the loss of the potential energy from the sodium intercalation. By subtracting this number from the AFM and FM energy difference, the stabilization due to the spin configurations can be found

| | $E_{Na2Mn3O7}$ | $E_{Na_{1.3}Mn3O7}$ | Energy difference ($E_{Na1.3}$ - $E_{Na2}$) | Stabilization due to spin ($E_{Na1.3, x}$ - $E_{Na2, x}$) - ($E_{Na1.3, non}$ - $E_{Na2, non}$) |
|---|---|---|---|---|
| non-magnetic | -105.8775411 eV | -99.26789 eV | 6.6096511 eV | 0 eV |
| AFM | -112.1268213 eV | -105.6659 eV | 6.4609213 eV | - 0.14897298 eV |
| FM | -112.119236 eV | -105.84291 eV | 6.276326 eV | -0.3333251 eV |

Table S4: Refined crystal structure of synthesized $Na_2Mn_3O_7$ (based on SSRL synchrotron XRD data and atomic coordinates refined). For refining the atomic coordinates of $Na_2Mn_3O_7$, the structural model for $Na_2Mn_3O_7$ was used as a starting point (ICDD PDF# 04-012-8515 [47]) for as-synthesized $Na_2Mn_3O_7$ and the transition metal and sodium ion atomic ratio adjusted to the synthetic ratio. The refinement procedure involved first refining the background, zero offset and lattice parameters and followed by the size and microstrain terms. This was followed by refining the atomic displacement parameters for all the elements.

| $Na_2Mn_3O_7$<br>Space Group: $P\bar{1}$ | a = 6.52625 Å, b = 6.95342 Å, c = 7.52371 Å<br>α = 109.106 °, β = 103.153 °, γ = 111.946 ° | | | | | | |
|---|---|---|---|---|---|---|---|
| No. | Atom | Label | x | y | z | Occ. | U |
| 1 | *Mn* | Mn1 | 0.07188 | 0.05889 | 0.22044 | 1 | 0.00253 |
| 2 | *Mn* | Mn2 | 0.37019 | 0.38275 | 0.08991 | 1 | 0.03995 |
| 3 | *Mn* | Mn3 | 0.22378 | 0.22834 | 0.63857 | 1 | 0.00254 |
| 4 | *Na* | Na1 | 0.23735 | 0.67737 | 0.50778 | 1 | 0.01900 |
| 5 | *Na* | Na2 | 0.30183 | 0.87018 | 0.09837 | 1 | 0.23450 |
| 6 | *O* | O1 | 0.28812 | 0.26622 | 0.20575 | 1 | 0.07587 |
| 7 | *O* | O2 | 0.17718 | 0.10517 | 0.77520 | 1 | 0.00747 |
| 8 | *O* | O3 | 0.04746 | 0.14838 | 0.00953 | 1 | 0.06791 |
| 9 | *O* | O4 | 0.15854 | 0.33936 | 0.44340 | 1 | 0.02092 |
| 10 | *O* | O5 | 0.71369 | 0.47284 | 0.10414 | 1 | 0.04802 |
| 11 | *O* | O6 | 0.12966 | 0.92510 | 0.43423 | 1 | 0.01437 |
| 12 | *O* | O7 | 0.42074 | 0.64950 | 0.32593 | 1 | 0.18439 |

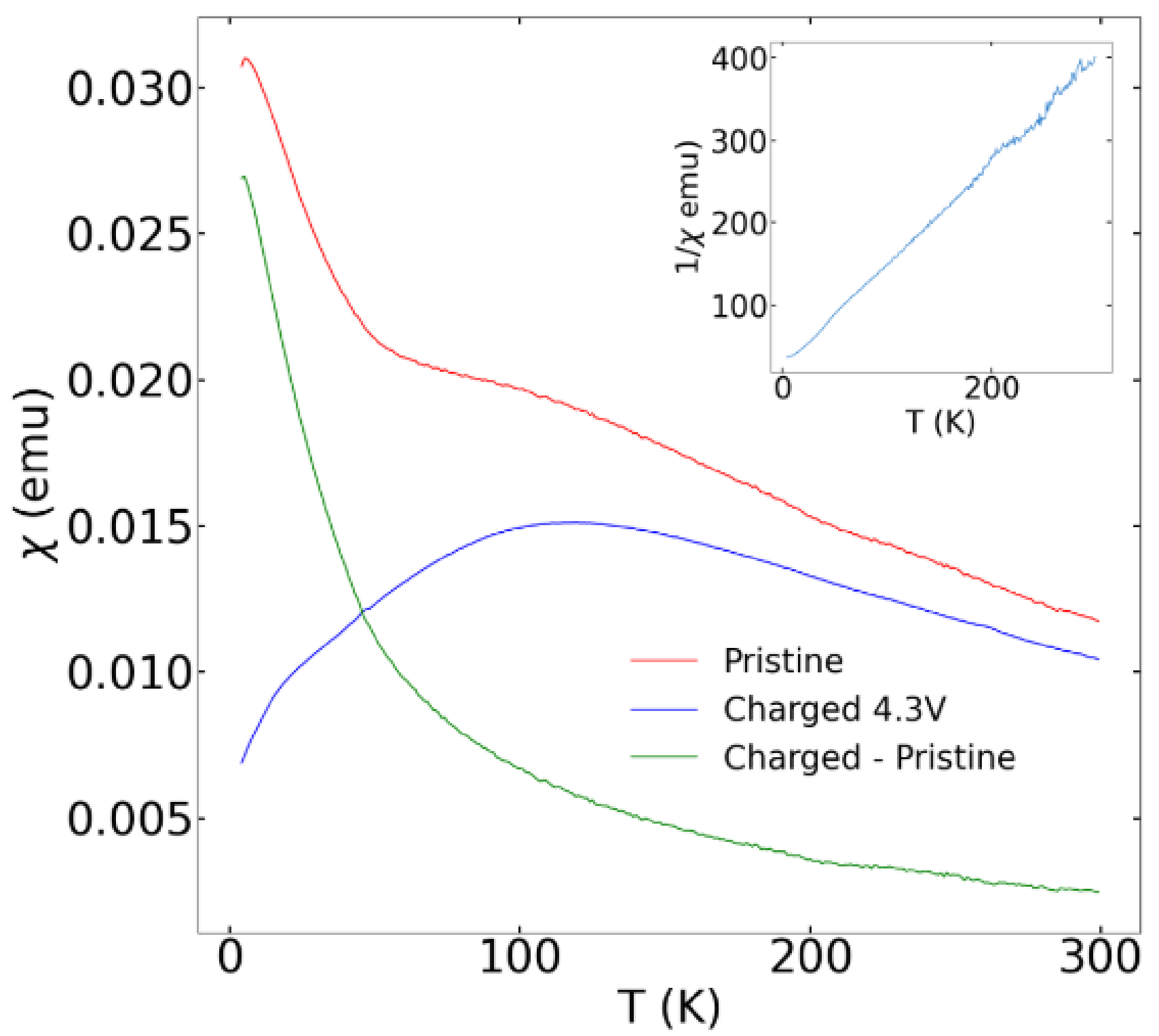

Figure S4: $\chi$ vs T for pristine (blue) and the deintercalated (red) sample along with a direct subtraction of the AFM pristine signal from the deintercalated signal (green). The inset shows the linear inverse $\chi$ of the subtracted signal, showing paramagnetic behavior

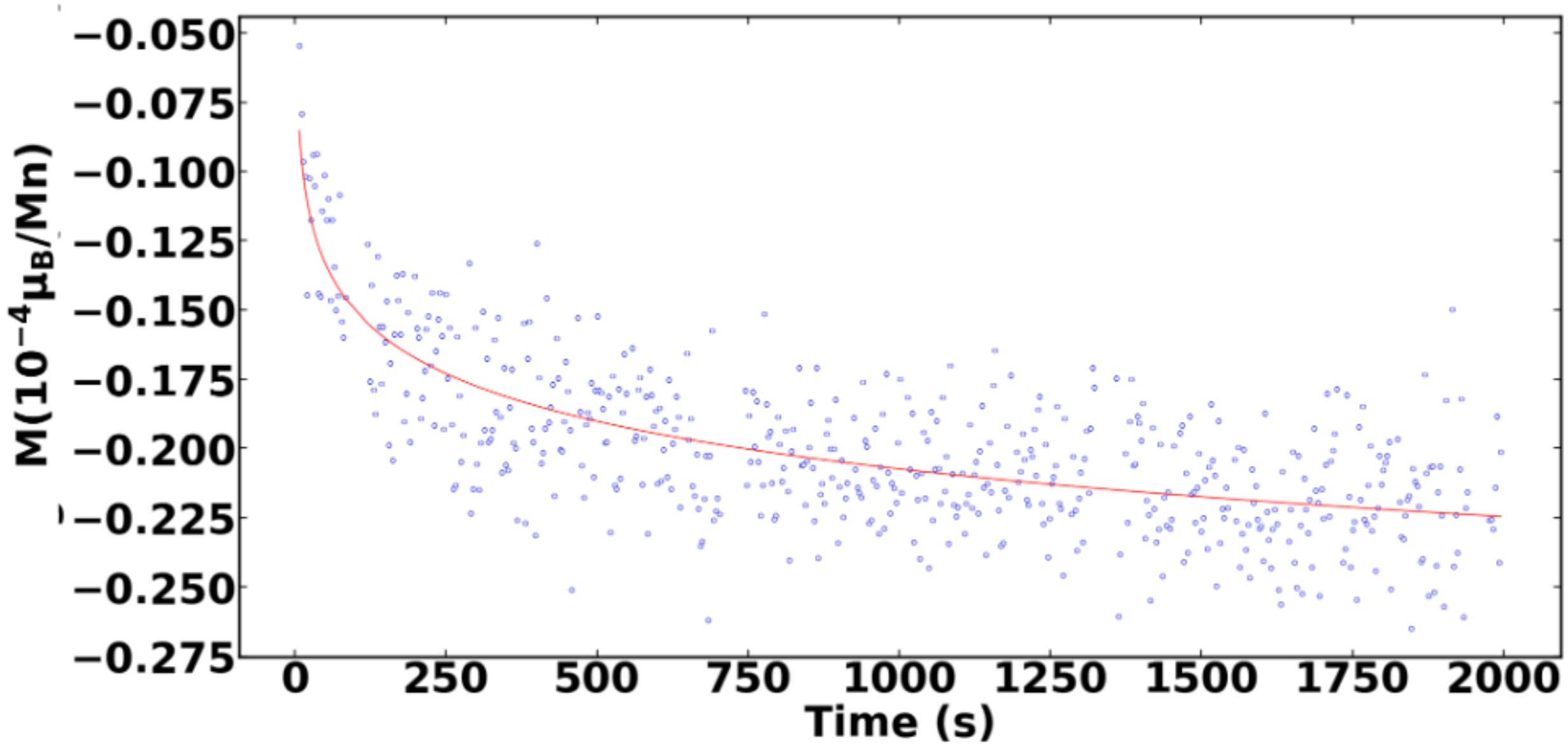

Figure S5. Thermoremanent magnetization (TRM) measurement after $Na_2Mn_3O_7$ was field-cooled in a field of 1000 Oe at 5K. The relaxations were measured after the magnetic field was removed. The remnant magnetization was on the level of system noise with high standard deviations and negative magnitude, potentially due to negative trapped fields [43].

Table S5. Parameters from fitting the time-dependent remnant magnetization to a modified stretched exponential function for $Na_2Mn_3O_7$.

| TRM Values | $M_0$ ($\mu_B/Mn$) | $M_g$ ($\mu_B/Mn$) | $\tau$ (s) | $n$ |
| --- | --- | --- | --- | --- |
| Absolute | -2.444e-05 | -1.609e-05 | 304 | 0.650 |
| St. Deviation | 3.029e-06 | 7.422e-06 | 166 | 0.189 |

**Notes and references**

1. Zhang, B., Lu, P., Tabrizian, R., Feng, P. X.-L. & Wu, Y. 2D Magnetic heterostructures: spintronics and quantum future. *Npj Spintron.* **2**, 6 (2024).
2. Sivý, D., Karl'ová, K. & Strečka, J. Quantum magnetism in Fe2Cu2 polymeric branched chains: insights from exactly solved Ising-Heisenberg model. *Front. Phys.* **12**, (2024).
3. Chandrasekharan, N. & Vasudevan, S. Dilution of a layered antiferromagnet: Magnetism in Mn x Zn 1 − x P S 3. *Phys. Rev. B* **54**, 14903–14906 (1996).
4. Upreti, D. *et al.* Medium-entropy Engineering of magnetism in layered antiferromagnet CuxNi2(1-x)CrxP2S6. Preprint at http://arxiv.org/abs/2408.02858 (2024).
5. Verzhbitskiy, I. & Eda, G. Electrostatic control of magnetism: Emergent opportunities with van der Waals materials. *Appl. Phys. Lett.* **121**, 060501 (2022).
6. Wu, Y., Li, D., Wu, C.-L., Hwang, H. Y. & Cui, Y. Electrostatic gating and intercalation in 2D materials. *Nat. Rev. Mater.* **8**, 41–53 (2023).
7. Schlottmann, P. Nematic and antiferromagnetic order in iron-pnictide superconductors. *Eur. Phys. J. B* **96**, 58 (2023).
8. Dimri, M. C., Khanduri, H. & Stern, R. Effects of aliovalent dopants in LaMnO3: Magnetic, structural and transport properties. *J. Magn. Magn. Mater.* **536**, 168111 (2021).
9. Dagotto, E. Complexity in Strongly Correlated Electronic Systems. *Science* **309**, 257–262 (2005).
10. Kajale, S. N. *et al.* Current-induced switching of a van der Waals ferromagnet at room temperature. *Nat. Commun.* **15**, 1485 (2024).
11. Molinari, A., Hahn, H. & Kruk, R. Voltage-Control of Magnetism in All-Solid-State and Solid/Liquid Magnetoelectric Composites. *Adv. Mater.* **31**, 1806662 (2019).
12. Sood, A. *et al.* Electrochemical ion insertion from the atomic to the device scale. *Nat. Rev. Mater.* **6**, 847–867 (2021).
13. Upreti, D. *et al.* Tuning Magnetism in Ising-type van der Waals Magnet FePS3 by Lithium Intercalation. Preprint at http://arxiv.org/abs/2407.12662 (2024).

14. The intercalation reaction of pyridine with manganese thiophosphate, MnPS3 | Journal of the American Chemical Society. https://pubs.acs.org/doi/10.1021/ja00046a027.

15. Aliev, A. *et al.* Two-Dimensional Antiferromagnetism in the [Mn3+xO7][Bi4O4.5−y] Compound with a Maple-Leaf Lattice. *Angew. Chem. Int. Ed.* **51**, 9393–9397 (2012).

16. Farnell, D. J. J. *et al.* Interplay between lattice topology, frustration, and spin quantum number in quantum antiferromagnets on Archimedean lattices. *Phys. Rev. B* **98**, 224402 (2018).

17. Electronic Frustration on a Triangular Lattice. https://www.science.org/doi/10.1126/science.1099387 doi:10.1126/science.1099387.

18. Haraguchi, Y., Matsuo, A., Kindo, K. & Hiroi, Z. Frustrated magnetism of the maple-leaf-lattice antiferromagnet MgMn 3 O 7 · 3 H 2 O. *Phys. Rev. B* **98**, 064412 (2018).

19. Haraguchi, Y., Matsuo, A., Kindo, K. & Hiroi, Z. Quantum antiferromagnet bluebellite comprising a maple-leaf lattice made of spin- 1 / 2 Cu 2 + ions. *Phys. Rev. B* **104**, 174439 (2021).

20. Fennell, T., Piatek, J. O., Stephenson, R. A., Nilsen, G. & Ronnow, H. M. Spangolite: an s=1/2 maple leaf lattice antiferromagnet? *J. Phys. Condens. Matter* **23**, 164201 (2011).

21. Schmoll, P., Jeschke, H. O. & Iqbal, Y. Tensor network analysis of the maple-leaf antiferromagnet spangolite. *arXiv e-prints* Preprint at https://doi.org/10.48550/arXiv.2404.14905 (2024).

22. Ghosh, P., Müller, T., Iqbal, Y., Thomale, R. & Jeschke, H. O. Effective spin-1 breathing kagome Hamiltonian induced by the exchange hierarchy in the maple leaf mineral bluebellite. *Phys. Rev. B* **110**, 094406 (2024).

23. Okuma, R., Yajima, T., Nishio-Hamane, D., Okubo, T. & Hiroi, Z. Weak ferromagnetic order breaking the threefold rotational symmetry of the underlying kagome lattice in \$\mathrm{CdC}{\mathrm{u}}_{3}{(\mathrm{OH})}_{6}{({\mathrm{NO}}_{3})}_{2}\ifmmode\cdot\else\textperiodcentered\fi{}{\mathrm{H}}_{2}\mathrm{O}\$. *Phys. Rev. B* **95**, 094427 (2017).

24. Pati, S. K. & Rao, C. N. R. Kagome network compounds and their novel magnetic properties. *Chem. Commun.* 4683–4693 (2008) doi:10.1039/B807207H.

25. Bhandari, H. *et al.* Magnetism and fermiology of kagome magnet YMn6Sn4Ge2. *Npj Quantum*

*Mater.* **9**, 1–10 (2024).

26. Guguchia, Z. *et al.* Hidden magnetism uncovered in a charge ordered bilayer kagome material $ScV_6Sn_6$. *Nat. Commun.* **14**, 7796 (2023).

27. Lee, Y. *et al.* Interplay between magnetism and band topology in the kagome magnets $R{\mathrm{Mn}}_{6}{\mathrm{Sn}}_{6}$. *Phys. Rev. B* **108**, 045132 (2023).

28. Luo, L. B. *et al.* Spin-glass behavior in hexagonal ${\mathrm{Na}}_{0.70}\mathrm{Mn}{\mathrm{O}}_{2}$. *Phys. Rev. B* **75**, 125115 (2007).

29. Vidya, R., Ravindran, P., Vajeeston, P., Kjekshus, A. & Fjellvåg, H. Effect of oxygen stoichiometry on spin, charge, and orbital ordering in manganites. *Phys. Rev. B* **69**, 092405 (2004).

30. V. Bazuev, G. *et al.* The effect of manganese oxidation state on antiferromagnetic order in $SrMn_{1-x}Sb_xO_3$ ($0 < x < 0.5$) perovskite solid solutions. *J. Mater. Chem. C* **7**, 2085–2095 (2019).

31. Tanasescu, S., Marinescu, C., Maxim, F., Sofronia, A. & Totir, N. Evaluation of manganese and oxygen content in $La_{0.7}Sr_{0.3}MnO_{3-\delta}$and correlation with the thermodynamic data. *J. Solid State Electrochem.* **15**, 189–196 (2011).

32. Dwivedi, G. D. *et al.* Switching of dominant magnetic exchange interactions between tetrahedral–octahedral and octahedral–octahedral sites in $(Mn_{1-x}Cr_x)_3O_4$ spinels. *J. Mater. Chem. C* **11**, 11312–11324 (2023).

33. Khmelevskyi, S., Ruban, A. V. & Mohn, P. Magnetic ordering and exchange interactions in structural modifications of $Mn_3Ga$ alloys: Interplay of frustration, atomic order, and off-stoichiometry. *Phys. Rev. B* **93**, 184404 (2016).

34. Abate, I. I. *et al.* Coulombically-stabilized oxygen hole polarons enable fully reversible oxygen redox. *Energy Environ. Sci.* **14**, 4858–4867 (2021).

35. Mortemard de Boisse, B. *et al.* Highly Reversible Oxygen-Redox Chemistry at 4.1 V in $Na_{4/7-x}[\square_{1/7}Mn_{6/7}]O_2$ (□: Mn Vacancy). *Adv. Energy Mater.* **8**, 1800409 (2018).

36. Venkatesh, C. *et al.* Magnetic properties of the one-dimensional $S=\frac{3}{2}$ Heisenberg antiferromagnetic spin-chain compound

${\mathrm{Na}}_{2}{\mathrm{Mn}}_{3}{\mathrm{O}}_{7}$. *Phys. Rev. B* **101**, 184429 (2020).

37. Saha, B., Bera, A. K., Yusuf, S. M. & Hoser, A. Two-dimensional short-range spin-spin correlations in the layered spin- 3 2 maple leaf lattice antiferromagnet Na 2 Mn 3 O 7 with crystal stacking disorder. *Phys. Rev. B* **107**, 064419 (2023).
38. Sonnenschein, J. *et al.* Candidate quantum spin liquids on the maple-leaf lattice. *Phys. Rev. B* **110**, (2024).
39. Maniv, E. *et al.* Antiferromagnetic switching driven by the collective dynamics of a coexisting spin glass. *Sci. Adv.* **7**, eabd8452 (2021).
40. Lee, C. C. *et al.* Superconducting to spin-glass state transformation in $\ensuremath{\beta}$-pyrochlore ${\mathrm{K}}_{x}$Os${}_{2}$O${}_{6}$. *Phys. Rev. B* **83**, 104503 (2011).
41. Spin Glasses | An Experimental Introduction | J A Mydosh | Taylor & Fr. https://www.taylorfrancis.com/books/mono/10.1201/9781482295191/spin-glasses-mydosh.
42. Coelho, A. A. TOPAS and TOPAS-Academic: an optimization program integrating computer algebra and crystallographic objects written in C++. *J. Appl. Crystallogr.* **51**, 210–218 (2018).
43. Kumar, N. & Sundaresan, A. On the observation of negative magnetization under zero-field-cooled process. *Solid State Commun.* **150**, 1162–1164 (2010).
44. Blöchl, P. E. Projector augmented-wave method. *Phys. Rev. B* **50**, 17953–17979 (1994).
45. Kresse, G. & Furthmüller, J. Efficient iterative schemes for ab initio total-energy calculations using a plane-wave basis set. *Phys. Rev. B* **54**, 11169–11186 (1996).
46. Heyd, J., Scuseria, G. E. & Ernzerhof, M. Hybrid functionals based on a screened Coulomb potential. *J. Chem. Phys.* **118**, 8207–8215 (2003).
47. Song, B. *et al.* Understanding the Low-Voltage Hysteresis of Anionic Redox in Na2Mn3O7. *Chem. Mater.* **31**, 3756–3765 (2019).
48. Raekelboom, E. A., Hector, A. L., Owen, J., Vitins, G. & Weller, M. T. Syntheses, Structures, and Preliminary Electrochemistry of the Layered Lithium and Sodium Manganese(IV) Oxides, A2Mn3O7. *Chem. Mater.* **13**, 4618–4623 (2001).

49. Chang, F. M. & Jansen, M. Darstellung und Kristallstruktur von Na2Mn3O7. *Z. Für Anorg. Allg. Chem.* **531**, 177–182 (1985).

50. Stephens, P. W. Phenomenological model of anisotropic peak broadening in powder diffraction. *J. Appl. Crystallogr.* **32**, 281–289 (1999).

51. Makuta, R. & Hotta, C. Dimensional reduction in quantum spin-$\frac{1}{2}$ system on a $\frac{1}{7}$-depleted triangular lattice. *Phys. Rev. B* **104**, 224415 (2021).

52. Delmas, C., Le Flem, G., Fouassier, C. & Hagenmuller, P. Etude comparative des proprietes magnetiques des oxydes lamellaires *A*CrO2 (*A* = Li, Na, K)—II. *J. Phys. Chem. Solids* **39**, 55–57 (1978).

53. Law, J. M., Benner, H. & Kremer, R. K. Padé approximations for the magnetic susceptibilities of Heisenberg antiferromagnetic spin chains for various spin values. *J. Phys. Condens. Matter* **25**, 065601 (2013).

54. Almeida, J. R. L. de & Thouless, D. J. Stability of the Sherrington-Kirkpatrick solution of a spin glass model. *J. Phys. Math. Gen.* **11**, 983 (1978).

55. Kanamori, J. Superexchange interaction and symmetry properties of electron orbitals. *J. Phys. Chem. Solids* **10**, 87–98 (1959).

56. Anderson, P. W. Antiferromagnetism. Theory of Superexchange Interaction. *Phys. Rev.* **79**, 350–356 (1950).

57. Nordblad, P., Svedlindh, P., Lundgren, L. & Sandlund, L. Time decay of the remanent magnetization in a CuMn spin glass. *Phys. Rev. B* **33**, 645–648 (1986).

58. Moriya, T. & Takahashi, Y. Spin Fluctuation Theory of Itinerant Electron Ferromagnetism —A Unified Picture. *J. Phys. Soc. Jpn.* **45**, 397–408 (1978).

59. Zhu, X., Edström, A. & Ederer, C. Magnetic exchange interactions in ${\mathrm{SrMnO}}_{3}$. *Phys. Rev. B* **101**, 064401 (2020).

60. Balents, L. Spin liquids in frustrated magnets. *Nature* **464**, 199–208 (2010).

61. Dong, Q. Y., Shen, B. G., Chen, J., Shen, J. & Sun, J. R. Spin-glass behavior and magnetocaloric

effect in melt-spun TbCuAl alloys. *Solid State Commun.* **151**, 112–115 (2011).

62. Sharma, M. K. & Mukherjee, K. Evidence of large magnetic cooling power and double glass transition in Tb5Pd2. *J. Magn. Magn. Mater.* **466**, 317–322 (2018).

63. Kumar, A. & Pandey, D. Study of magnetic relaxation, memory and rejuvenation effects in the cluster spin-glass phase of *B*-site disordered Ca(Fe1/2Nb1/2)O3 perovskite: Experimental evidence for hierarchical model. *J. Magn. Magn. Mater.* **511**, 166964 (2020).

64. Fita, I. *et al.* Competing exchange bias and field-induced ferromagnetism in La-doped $\mathrm{BaFe}{\mathrm{O}}_{3}$. *Phys. Rev. B* **95**, 134428 (2017).

65. Morales, R. *et al.* Exchange-Bias Phenomenon: The Role of the Ferromagnetic Spin Structure. *Phys. Rev. Lett.* **114**, 097202 (2015).

66. He, J. Q. *et al.* Competing two-phase coexistence in doped manganites: Direct observations by in situ Lorentz electron microscopy. *Phys. Rev. B* **82**, 224404 (2010).

67. Zhou, W. *et al.* Kondo behavior and metamagnetic phase transition in the heavy-fermion compound ${\mathbf{CeBi}}_{\mathbf{2}}$. *Phys. Rev. B* **97**, 195120 (2018).

68. Zhang, M., Kamavarum, V. & Reddy, R. G. New electrolytes for aluminum production: Ionic liquids. *JOM* **55**, 54–57 (2003).